\documentclass[prl,onecolumn,superscriptaddress,nofootinbib]{revtex4-1}
\usepackage{graphicx}
\usepackage{color}
\usepackage{amsmath}
\usepackage{amsfonts}
\usepackage{mathrsfs}
\begin{document}

\title[Phyllotaxis, topological evolution]{Phyllotaxis: a framework for foam topological evolution}

\author{ Nicolas Rivier }
\email{rivier@ipcms.unistra.fr}
\affiliation{ Institut de Physique et Chimie des Mat\'eriaux de Strasbourg (IPCMS), Universit\'e de Strasbourg,
F-67084 Strasbourg, cedex }

\author{Jean-Fran\c cois Sadoc}
\email{jean-francois.sadoc@u-psud.fr}
\affiliation{Laboratoire de Physique des Solides (CNRS-UMR 8502), B{\^a}t. 510, Universit{\'e} Paris-sud, F 91405 Orsay cedex}

\author{Jean Charvolin}
\affiliation{Laboratoire de Physique des Solides (CNRS-UMR 8502), B{\^a}t. 510, Universit{\'e} Paris-sud, F 91405 Orsay cedex}

\begin{abstract}
Phyllotaxis describes the arrangement of florets, scales or leaves in composite flowers or plants (daisy, aster, sunflower, pinecone, pineapple). As a structure, it is a geometrical foam, the most homogeneous and densest covering of a large disk by Voronoi cells (the florets), constructed by a simple algorithm: Points placed regularly on a generative spiral constitute a spiral lattice, and phyllotaxis is the tiling by the Voronoi cells of the spiral lattice. Locally, neighboring cells are organized as three whorls or parastichies, labelled with successive Fibonacci numbers. The structure is encoded as the sequence of the shapes (number of sides) of the successive Voronoi cells on the generative spiral. We show that sequence and organization are independent of the position of the initial point on the generative spiral, that is invariant under disappearance ($T2$) of the first Voronoi cell or, conversely, under creation of a first cell, that is under growth. This independence shows how a foam is able to respond to a shear stress, notably through grain boundaries that are layers of square cells slightly truncated into heptagons, pentagons and hexagons, meeting at four-corner vertices, critical points of $T1$ elementary topological transformations. \\
Preprint of a paper published in The European Physical Journal E, 39(1), 1-11; DOI 10.1140/epje/i2016-16007-8
\end{abstract}

\maketitle

\section{Introduction}
A foam can be represented as a space-filling cellular network, a tessellation, usually random. In 3D, the cells are bounded by interfaces connected three by three on edges that meet four by four on vertices. In 2D, the cells are bounded by edges that meet three by three on vertices. This generic description holds unless the meeting point has been especially adjusted.
The foam has a topological dual in which a vertex is associated with each cell and an edge joining two vertices traverses
the interface between two cells. Generically, the cells in the dual are simplices, triangles in 2D and tetrahedra in 3D.
It is often assumed
that the elements of the foam are physical elements, i.e the cells are closed bubbles filled by a discontinuous gas phase and the space between them (the interface) is filled by the continuous liquid phase. Edges and vertices, called Plateau borders are distinct in dry foams \cite{weaire}. In phyllotaxis, the interfaces between florets are not physical entities. Nevertheless, topological defects such as dislocations and grain boundaries are defined precisely and their evolution under stress can be followed.

\begin{figure}[tbp]
\resizebox{0.9\textwidth}{!}{%
\includegraphics{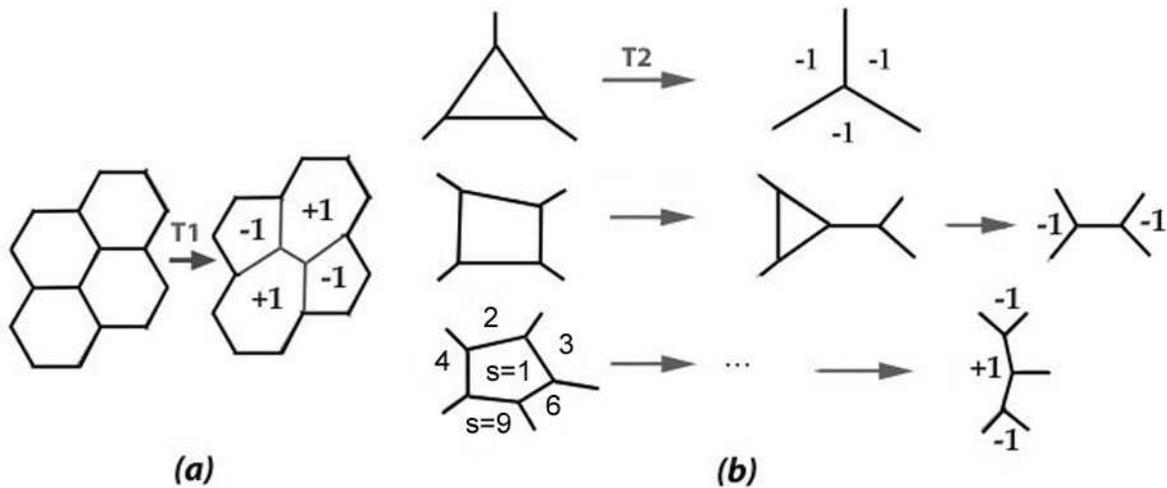}
}
\caption{a) $T1$: Shear flips the interface between two cells, switches neighbors and produces  a pair of dislocations $5/7\setminus7/5$. b) $T2$: Cell disappearance. On the left are three examples of the disappearing cell with $m=3,4,5$ sides. On the right is the trace of the disappearing cell left on the number of sides of its neighbors. The trace is topologically unique for $m \leq 5$. The $T2$ follows one or two $T1$ for $m=4$ or  $m=5$, respectively \cite{dubertret}.
Since we are concerned in this paper with the disappearance of pentagonal cells in the core of phyllotaxis,  cells for $m=5$ are numbered with $s=1$ as the first cell on the generative spiral. Its neighbors are, clockwise, $s=2, 3, 6, 9 \textrm{~and~} 4$, on parastichies $1, 2, 5, 8  \textrm{~and~} 3$. The pentagonal cell $s=1$ disappears by removal of the interface $1|4$; cells $s=2$ and $s=9$ lose a side, and cell $s=4$ gains one.}

\label{fg1}
% %

\end{figure}

This representation of the foam as a space-filling cellular complex simplifies greatly the description of its deformation and of its evolution. The foam is in  quasi-static, statistical equilibrium \cite{dubertret,dubertret2}, punctuated by elementary, local topological transformations involving an {\it interface}. As shown on the figure \ref{fg1},  the flip of an interface (usually called $T1$ transformation)  produces
a switch of neighbor cells.
The disappearance of an interface (a coalescence event) leads to the disappearance of one of the two cells that it separates (called $T2$ process, although $T2$ is an elementary transformation for three-sided cells only \cite{dubertret}).
These transformations introduce
topological defects, notably dislocations, with specific and dramatic effects on the properties of the material, not least on its mechanical resistance or plasticity \cite{mott,miri}. Topological defects can be described for two dimensional foams (a packing of polygonal cells) or three dimensional foams.  Dislocations have to be referred to a perfect structure which for 2D foams is generally a hexagonal structure \cite{miri}. In three dimensional structures the reference order is not so simple to define and $T1$ transformations are not unique. In this paper  we are more concerned by 2D foams, even if some points can be generalized. Then, on a hexagonal reference order, dislocations are dipoles formed by a pentagonal and a heptagonal cell. A $T1$ transformation in a hexagonal structure generates a quadrupole of two such dipoles .

% %______________________Fig.1___________________________________ %

%________________________________________________________________ % % %
%

\begin{figure}[tbp]
\resizebox{0.9\textwidth}{!}{%
\includegraphics{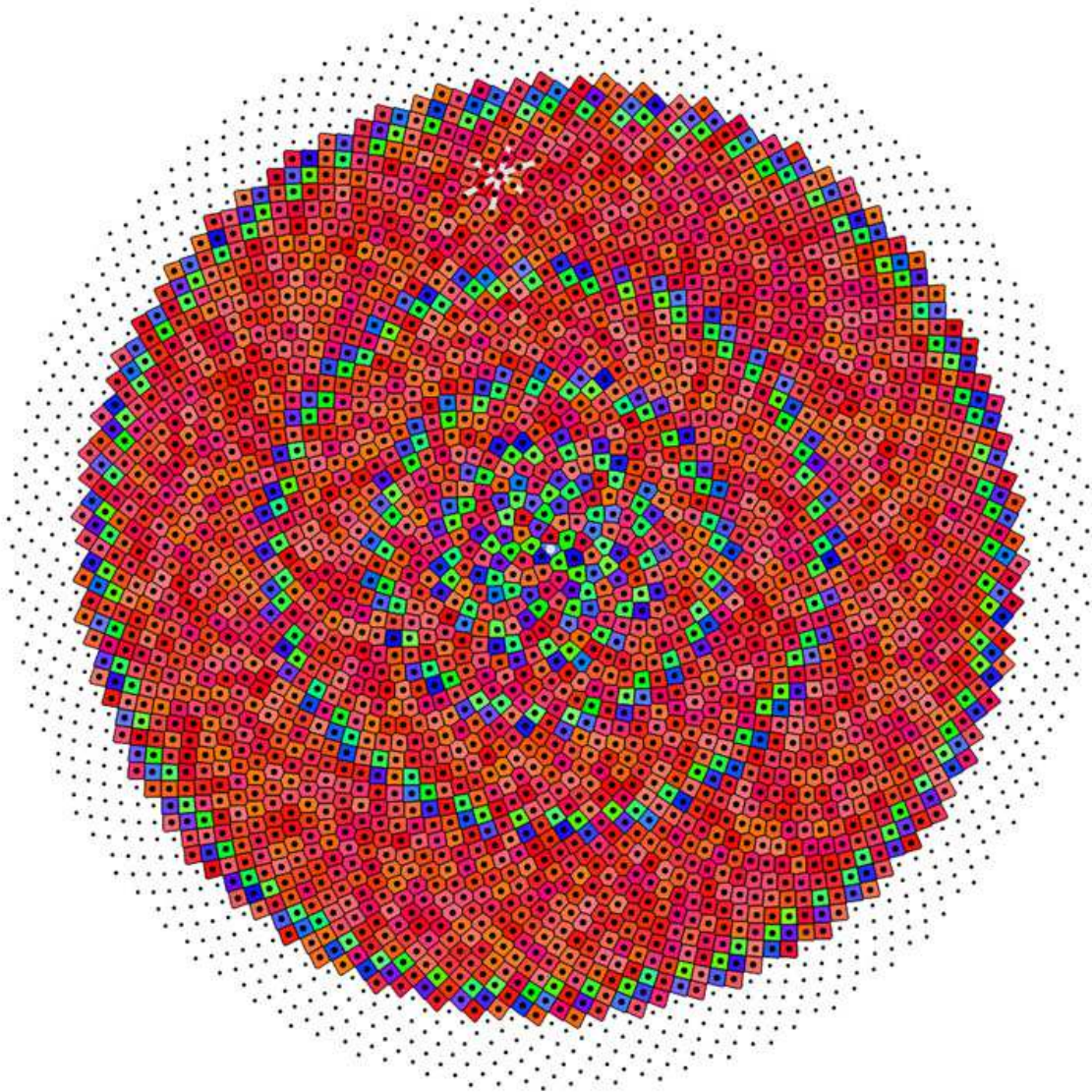}
}
\caption{ Planar phyllotaxis ($n=3000$ points organized on the plane according to the algorithm of phyllotaxis with the golden ratio and uniform density). Each point is surrounded by its polygonal Voronoi cell with a number of sides equal to the number of first neighbors around the point. Dark, medium grey and white cells are respectively pentagons, hexagons and heptagons. The three visible spirals of first neighbors around each hexagonal cell (white lines) are called parastichies. The white dot marks the origin point $s=0$ which is also the origin of the generative spiral \cite{sadoc2}.}
\label{fg2}
% %

\end{figure}

An array of dislocations constitutes  a grain boundary, with its own, remarkable dynamics.  Grain boundaries are easily seen in the bubble rafts of Bragg and Nye \cite{mott}, separating crystalline hexagonal grains of different orientations.
We shall see that grain boundaries occur in phyllotaxis (Fig.\ref{fg2}) where they play an essential part in the growth of the structure.

The representation of a foam as a space-filling cellular network is a simplification of the complex, biphasic material where an interface is a concrete  physical object (the continuous fluid phase). But it is also a natural extension when the interfaces have no concrete existence per se, as in granular packings or in phyllotaxis where the constituents are individual grains or florets packed together.

Phyllotaxis (leaf arrangement in botanics - e.g. Fig.\ref{fg2}) is a space-filling structure with specific symmetries different from the translation invariance of a crystal, the inflation-deflation symmetry of a quasicrystal and the global uniformity of a random foam. Like a crystal, it is based on a lattice, but it is a spiral lattice, and the structure is encoded as a one-dimensional sequence of all the cells in the pattern. Like any foam,
phyllotaxis is structured into layers \cite{oguey,AsBoRi}.

% %______________________Fig.2___________________________________ %

%________________________________________________________________ % % %
%
\section{The tiling of phyllotaxis}
\subsection{A spiral lattice}
Phyllotaxis describes the arrangement of florets, scales or leaves in composite flowers or plants (daisy, aster, sunflower, pine cone, pineapple). Geometrically, it is a foam (albeit ordered and organized), the most homogeneous and densest covering of a large disk by Voronoi cells (the florets) \cite{ridley}.  The Voronoi cell (or Dirichlet domain) associated with a point is defined as the region of space nearer to it than to any other point in the set. Points placed regularly on a generative spiral constitute a spiral lattice, and phyllotaxis is the tiling by the Voronoi cells of the spiral lattice. The number of points on the generative spiral increases as the flower grows.

First, one constructs on a surface of constant Gauss curvature (plane, sphere, cone, cylinder) a spiral lattice defined by a parametric equation (with a parameter $s$) for a generative spiral. Discrete points, dual of cells,  are set regularly at integer values of $s = 0,1,2,...,n$.
Changing the curvature can have a real meaning, as for pine cones or pineapples where the physical substrate is not planar, but it can also be a way of taking into account local changes in the sizes of the cells, such as larger florets on the periphery of sunflowers.

Then, one associates to each point a Voronoi cell or tile (Dirichlet region) \cite{sadoc1,sadoc2}. The position of each point $s$ is given by its polar coordinates $(\rho,\theta)$ for example, on the plane
\begin{eqnarray}
\rho(s)=a \sqrt{s} \textrm{~and~}   \theta(s)=(2 \pi/ \tau) s
\end{eqnarray}
(here, Fermat's spiral $\rho \sim \sqrt{\theta}$
is the generative spiral)
and the azimuthal angle between two successive points on the spiral is $2\pi/\tau$, where $\tau = (1+\sqrt{5})/2$ is the golden ratio
%_______________________footnote1_________________________________________
\footnote{The golden ratio implies that all transitions between parastichies at grain boundaries are ``regular'' \cite{koch1}: The concentric defect circles have all a self-similar structure, with alternating chirality, and the cells are almost square. Any noble irrational would yield the same result, but only after some stage, and the core would be slightly different. Any other quadratic irrational encompasses ``singular'' transitions between parastichies at grain boundaries with a structure marked by intermediate convergents \cite{koch1,rivier2,rivier3}, and with rectangular cells (rather than square).}.
%_____________________________________________________

We take this azimuthal law as part of the model leading to the lowest anisotropy of the cells and to the densest packing under these conditions. (A hexagonal structure is the densest 2D packing solution for an infinite crystal or for finite crystals with fitting boundaries. Phyllotaxis is the densest packing for circular boundaries or issued from an isotropic seed. As the structure grows by increasing the number of points on the generative spiral, the boundary remains close to circular. This optimal compactness explains why we have used phyllotaxis as a model for a 2D foam confined inside a circular boundary \cite{sadoc1}.  The phyllotactic structure is also hyperuniform \cite{torquato}
: although ordered, it is under steady strain through growth, but remains invariant under the elementary, local topological transformations ($T1$ and $T2$) that the strain induces.)

The phyllotactic pattern of a set of 3000 points on a plane
with their Voronoi cells is shown in figure~\ref{fg2}.  The structure is defined by the generative spiral which is not visible: the cells corresponding to successive points $s = 0,1,2,...,n$ on the generative spiral are not neighbors.  But the pattern does exhibit visible spirals of nearest neighbor cells. Through every hexagonal cell there are three such spirals, called by botanists ``parastichies'': The neighbors
%_______________________footnote2_________________________________________
\footnote{Indeed  $\theta(s+f_i) \simeq \theta(s)$ after $f_{i-1}$ turns of the generative spiral. Here, the $f$'s are successive Fibonacci numbers ($f_{i}=f_{i-1}+f_{i-2}$ with $f_{0}=0$ and $f_{1}=1$) and $f_{i-1}/f_{i}$ are rational approximants to $1/ \tau$.}
%------------------
to cell $s$ are $s\pm f_{i-1},s\pm f_{i-2},s\pm f_{i-3}$.
The parastichies split up or merge on the non-hexagonal cells which are concentrated on thin rings of defects. On the defect rings, the two parastichies that continue from one grain to the next are orthogonal and the cells are truncated rectangles \cite{rivier3}.
The defects are pentagon-heptagon dipoles acting as dislocations, gathered into grain boundaries. The grain boundaries are the thin rings  separating  the grains, thicker annuli of hexagonal cells.
Each dislocation introduces a new parastichy to maintain the density of points or cells as constant as possible.
The underlying arithmetic of this concentric organization \cite{sadoc2} is that of Fibonacci numbers. It can be summarized as a sequence $c(s)$ of cell types \{hexagon, pentagon, heptagon\} along the generative spiral, that is an encoding of the structure to be discussed below
%_______________________footnote3_________________________________________
\footnote{For the core and the innermost grain boundary of Fig.\ref{fg2}, the sequence $c(s)$ of cell types is $5,5,6,7,7,7,6,5,5,6,[(6,6,6,6,6,7,7,7),(6,6,6,6,6),(5,5,5,5,5,5,5,5)]$, with the point corresponding to cell $s=0$ at the origin of the generative spiral. The numbering of the successive cells along the generative spiral $s=0-30$ is given explicitly in Fig.\ref{exfig9} (see Appendix 2). Parastichy $13$ begins already at cell $s=2$. The innermost grain boundary  $[(6,6,6,6,6,7,7,7),(6,6,6,6,6),(5,5,5,5,5,5,5,5)]$ has 21 cells $s=10-30$. The core can be cleared up by starting the numbering from $s=1$ (eliminating cell $s=0$ by removing the boundary $0|2$) and making two neighbor flips to remove the $13$ parastichy from cells $s=2,3$. The sequence of cell types is then $5,6,6,6,6,6,6,6,6,[(7,7,6,6,6,6,6,7),(6,6,6,6,6),(5,5,5,5,5,5,5,5)]$. The innermost grain boundary  (21 cells $s=10-30$) is
$[(7,7,6,6,6,6,6,7),(6,6,6,6,6),(5,5,5,5,5,5,5,5)]$. Details in Fig.\ref{newfig}.}.
%____________________________________________________________________
% %______________________ex Fig. 9 ref appendix______________________________ %
\begin{figure}[tbp]
\resizebox{0.4\textwidth}{!}{%
\includegraphics{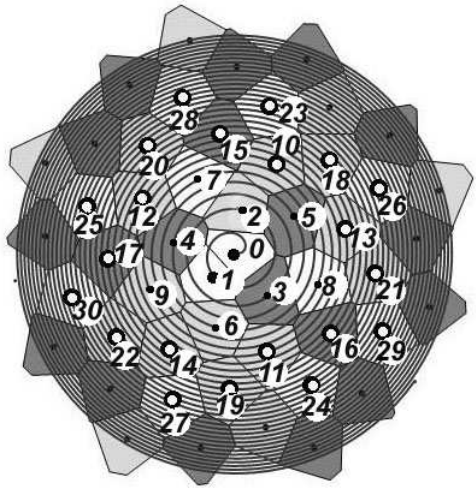}
}

\caption{Core of the phyllotactic structure of Fig.\ref{fg2}, with the generative spiral. It is Fig.\ref{newfig}a with cells $s=0-30$ numbered. Here, white cells are pentagons and dark grey, heptagons. The innermost defect ring has twenty one cells $s=10-30$ (open circles). The parastichies are spirals of neighbor cells. For instance, parastichies $5$ are the five spirals $s=2-7-12-17-22$, $3-8-13-18$, $4-9-14-19$, $5-10-15-20$ and $6-11-16-21$, with $3$ turns of the generative spirals between neighbor cells.}
\label{exfig9}
% %

\end{figure}
Incidentally,  the terminology ``spiral lattice'' is well-chosen
since the parastichies play the role of reticular lines in a classical 2D lattice.  We call {\it core} the open disk of cells inside the innermost  ring of defects.

\subsection{Plane or curved phyllotaxis}

Phyllotaxis can be viewed as a succession of bent hexagonal grains connected by grain boundaries, outside an inner core. Bending is a shear stress that can be related to the curvature of the substrate. On the plane, using a Fermat spiral as generative spiral with a radial law demanding equi-sized cells, the foam is subjected to a (Poisson) shear stress, i.e. compressed radially and expanded azimuthally. By contrast, a hexagonal crystal wrapped with chirality on a cylinder has no shear stress. This wrapped hexagonal crystal can be conformally mapped on a plane to yield a phyllotaxis that is a conformal (single) crystal with only hexagonal cells (outside a central core) and zero shear strain \cite{rothen}. The sizes of the florets increase radially outwards and the generative spiral is the equiangular spiral $\rho \sim \exp{\theta}$ (Bernoulli).
The curvature of the substrate, the nature of the generative spiral and the sizes of the cells are therefore intertwined properties of the phyllotactic structure.

A grain is denoted as $(f_{i-1},f_{i-2},f_{i-3})$ (i.e. hexagonal cells in phyllotaxis $(f_{i-1},f_{i-2},f_{i-3})$) and encoded as $...,6,6,6,...$ in the sequence $c(s)$.
Spherical phyllotaxis
%_____________________ footnote4 ________
\footnote{The algorithm used to build the phyllotactic pattern on the sphere has the position of point $s$
given by the spherical coordinates $(\rho,\phi,\theta)$ \cite{sadoc2}. Specifically, $\rho=R$ the sphere radius, $\theta= 2 \pi \lambda s$, the azimuthal angle and $\phi=-\arcsin(s^\prime/\nu)+\pi/2$, the polar angle.  The total number of points on the sphere  is $n=2\nu+1$ and the generative spiral is symmetric through mid-equator $s^\prime = 0$, and pinned at that point.
In cylindrical coordinates $(\rho,\theta, z), z= R \cos \phi $, the generative spiral $z/R = s^\prime/\nu$ is a regular (constant pitch) helix on a finite cylinder, a familiar representation of cylindrical phyllotaxis (pineapple) \cite{coxeter,koch1,koch2}. With this choice, the integer $s^\prime$ goes from $s^\prime=-\nu$ (south pole $z = -R, \phi=\pi$) to $s^\prime=\nu$ (north pole $z = R, \phi=0$) with $s^\prime=0$ for the point $(z = 0, \phi=\pi/2$) where the generative spiral crosses the equator.  For comparison with plane phyllotaxis, let point $s=0$ be the south pole using $s=s^\prime+\nu$.}
%______________________________________________________________
is obtained by the mapping of points on a finite cylinder tangent at the equator to the sphere with the same radius $R$ and height $2R$.
The finite cylinder and the sphere have the same area $4\pi R^2$.
Points on the cylinder are located on a regular helix defined by $\theta= (2 \pi/\tau) s$ and a constant pitch $h$ related to the number of points on the sphere. They constitute a perfect lattice without
defects wrapped on the cylinder, with a constant  density of points. Generically,  a point $s$ has six neighbors at positions $s\pm\delta s$ with the $\delta s$ equal to three successive Fibonacci numbers  depending on the pitch of the helix drawn on the cylinder. The Voronoi cells are then all hexagons with equal area.
For particular values of the pitch, the Voronoi cells are squares and the tessellation by squares has non-generic four-corner boundaries with degree $z = 4$ (the critical point of a $T1$ transformation)  \cite{coxeter},\cite{koch1}.

The points on the cylinder are mapped on the sphere orthogonally to the polar axis so that the area per point on the sphere is the same as on the cylinder (the area of a zone between two parallel planes is the same on a sphere of radius $R$ and on the tangent cylinder orthogonal to the planes).
But such a projection on the sphere, even if it conserves the area of the Voronoi cells
and the pitch of the helix, introduces inhomogeneous shear in the structure. This shearing, at constant area, is given by a compression factor $\sin(\phi)$ changing the length of the equator into that of a parallel (at $\phi$), and an expansion factor along meridians.
Shear can change the incidence relations between neighbors so that first neighbors are not necessarily the same as on the crystalline cylinder and defects (non-hexagonal cells) appear, notably on the polar circles (Fig.\ref{f6}) that separate topologically the cylinder from the two hemispherical polar caps.

Projection on a tangent cone is different because the generative spiral on the cone cannot have constant pitch and keep a constant density of points. $h(s)$ must vary linearly with $s$.

\subsection{Grain boundaries}
In plane phyllotaxis with a Fermat spiral and in spherical phyllotaxis,
cells are of equal area. But they are not all hexagonal: there are annular crystalline grains of hexagonal cells (traversed by three visible reticular lines in the form of spirals, the parastichies), separated by grain boundaries that are circles of dislocations ($d$: dipole pentagon/heptagon) and square-shaped topological hexagons ($t$: squares with two truncated adjacent vertices), in a quasiperiodic sequence $d t d d t d t \dots $. The two main parastichies cross at right angle through the grain boundaries. \cite{rivier2}.

The dislocations in the grain boundaries  induce a shear strain in the grains.
It is actually a Poisson shear, associated with radial compression between two circles of fixed, but different length. Thus, elastic and plastic shear can be readily absorbed by a polycrystalline phyllotactic structure described by several successive grain boundaries with
successive Fibonacci numbers. The packing efficiency problem is thereby solved: One grain boundary constitutes a perfect circular boundary for the disk into which objects are to be packed   (see also \cite{sadoc3}).
Grain boundaries can be moved (dislocation glide) by local neighbor exchanges ($T1$). In that way, the phyllotactic structure responds easily, locally and naturally to an external force that is expressed in the curvature of the substrate. A grain bounded by two boundaries of fixed length and quasicrystalline topology, is depleted as the curvature becomes less positive.

A circular ring of defects constitutes a grain boundary between grain $(f_{i-1},f_{i-2},f_{i-3})$ (hexagonal cells in phyllotaxis $(f_{i-1},f_{i-2},f_{i-3})$) and grain  $(f_{i},f_{i-1},f_{i-2})$.
It has a given length of $f_{i+1}$ non-hexagonal cells and uses $f_{i}$ turns of the generative spiral. The $f_{i+1}$ non-hexagonal cells are $f_{i-1}$ heptagons, $f_{i-2}$ hexagons (with the two smaller sides not opposite but next to adjacent) and $f_{i-1}$ pentagons, ordered in the quasiperiodic sequence $d t d d t d t \dots $. (Details in Fig.\ref{fg4}a or in Fig.A1 of \cite{sadoc2}:  strip cut in a square lattice). It is denoted as $[ f_{i-1},f_{i-2},f_{i-1}]$, and encoded as $[7,7, \dots,6,\dots, 5,5,\dots ]$ in the sequence $c(s)$.

On the grain boundary, the Voronoi cells are square-shaped, four incident on a vertex which is a critical point of a $T1$ (a natural four-corner boundary, with the two main parastichies crossing at right angle \cite{coxeter}). The squares are slightly truncated to specify the number of sides of the neighboring cells, and it is the truncation mode that flips ($T1$) under shear or growth. The hexagonal cells constituting a grain lie between the pentagons of one grain boundary and the heptagons of the next boundary.

The grain boundary in phyllotaxis is a circular ring of defects, the cells are squares, slightly truncated, and the main parastichies cross perpendicularly, so that the grain boundary can be represented as a strip in a square lattice, with the thin diagonal lines identified (Fig.\ref{fg4}a). The strip is mapped into a circular ring of defects (non-hexagonal cells) of Fig. 2. The mapping is conformal (it conserves the angles), so that the orthogonality of the main parastichies and the square shape of the cells are conserved.

In planar and spherical phyllotaxis, the innermost ring of defects contains five more pentagons (of topological charge $+1$) than heptagons (topological charge $-1$) \cite{miri}. This is because the topological charge of a disk (flattened hemisphere) must be $+6$, with $5$ taken up by the innermost {grain} boundary (polar circle) and $+1$ by the pole ($s=1$, a pentagonal cell). The core (polar cap) is the open disk of cells inside the innermost ring of defects (open circles in Fig.\ref{exfig9}).

Since the successive grain boundaries are circles of given length, they are better separated on a substrate with a more positive Gaussian curvature, and more crowded on a more negative curvature substrate.
Indeed, as we shall see later, spheres $n = 43-75$ have the simplest structure: only one grain $(13,8,5)$ separated from two polar caps of three cells each $(5,6,6)$ and $(6,6,5)$ by grain boundaries (polar circles) of thirteen cells each,\\
$[(6,6,6,6,6),(6,6,6),(5,5,5,5,5)]$ and $[(5,5,5,5,5),(6,6,6),(6,6,6,6,6)]$. (See Fig.\ref{f6}a). In smaller spheres $n = 16-42$, the grain is $(8,5,3)$ and the polar circle fills up the polar cap $[(5,6,6),(5,5,5,5,5)]$ with its $f_{6}= 8$ cells. The pole is the common vertex of cells $s=1,2,3$. Larger spheres are discussed in Appendix 1.

\section{Phyllotaxis as a foam}

The phyllotactic structure has global features, such as the circular grain boundaries, visible in Fig.\ref{fg2}, but also local features, the organization into parastichies, whorls of neighbor cells. Both features are invariant under the local, elementary topological transformations $T1$ and $T2$ that are driving cell disappearance in foam evolution (see Fig.1 and \cite{dubertret}). Moreover, the cells in grain boundaries, which are truncated squares, almost meet at four-corner vertices, the critical point  for $T1$. Thus, phyllotaxis is a topologically stable but malleable arrangement.

The grain boundaries (defect rings) constitute specific layers, a concept in (disordered) foams that  uses the notion of topological equi-distance \cite{oguey}.

\subsection{Layers and defect inclusions in foam}
\begin{figure}[tbp]
\resizebox{0.5\textwidth}{!}{%

\includegraphics{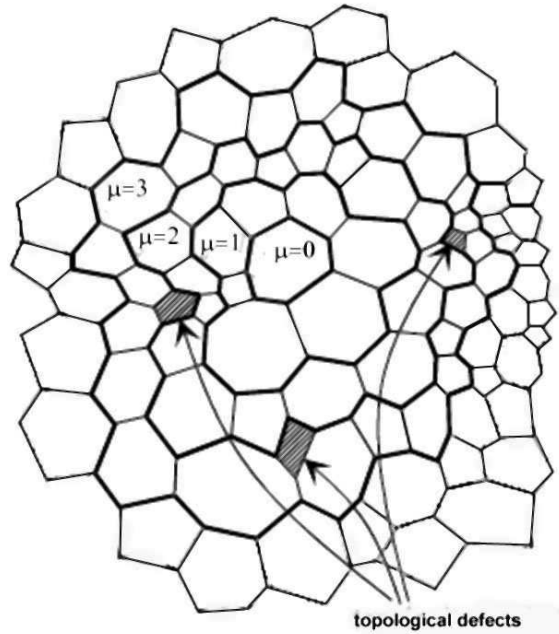}
}

\caption{ Example of successive layers in foam (from \cite{aste}). The nearest neighbors to the $7$-sided cell $\mu = 0$ are seven cells at topological distance $1$, constituting layer $\mu =1$. And so on, for layers $\mu = 2,3, \dots$. Note that three defect inclusions (one five- and two four-sided cells) appear (shaded) in layer $3$. They are at topological distance $3$ from the origin cell, neighbors to layer $2$ but not to any cell of layer $4$.
}
\label{fg3}

\end{figure}

Layers are best illustrated in Fig.\ref{fg3}, taken from \cite{aste}. The nearest neighbors to an
$n$-sided cell $\mu = 0$ are $n$ cells at distance $1$, thereby defining topological distance. They constitute layer $\mu =1$. And so on, for layers $\mu = 2,3, \dots$. Three defect inclusions (shaded) appear in layer $3$. They are neighbor cells to layer $2$ but not to any cell of layer $4$.

%______________________ Fig.3_new__________________________________

%

%________________________________________________________________

A foam (space-filling cellular complex) can be decomposed into successive layers
$\mu-1,\mu,\mu+1,\cdots$ going up (or outwards), or $\mu+1,\mu,\mu-1,\cdots$ going down (inwards). The cells of the layer are all at the same topological distance to an origin layer $\mu=1$, going up, or $\mu=\mu_{max}$ going down. \cite{oguey}.  In phyllotaxis, the layers are naturally anchored on the defect rings, so that the origin layer is specified only with respect to the nearest defect ring.
The topological distance between two cells is the minimum number of interfaces one has to cross to go from (the interior of) one to the other. Nearest neighbor cells are distant of $1$.  All cells of layer $\mu$ are nearest neighbors to at least one cell of layer $\mu-1$ going up ($\mu+1$ going down); most, but in general not all, are also neighbors to layer $\mu+1$ (resp. $\mu-1$). A cell without direct contact with layer $\mu+1$ (resp. $\mu-1$) is called defect inclusion.

%

% %______________________Fig.4___________________________________ %
\begin{figure}[tbp]
\resizebox{0.8\textwidth}{!}{%
\includegraphics{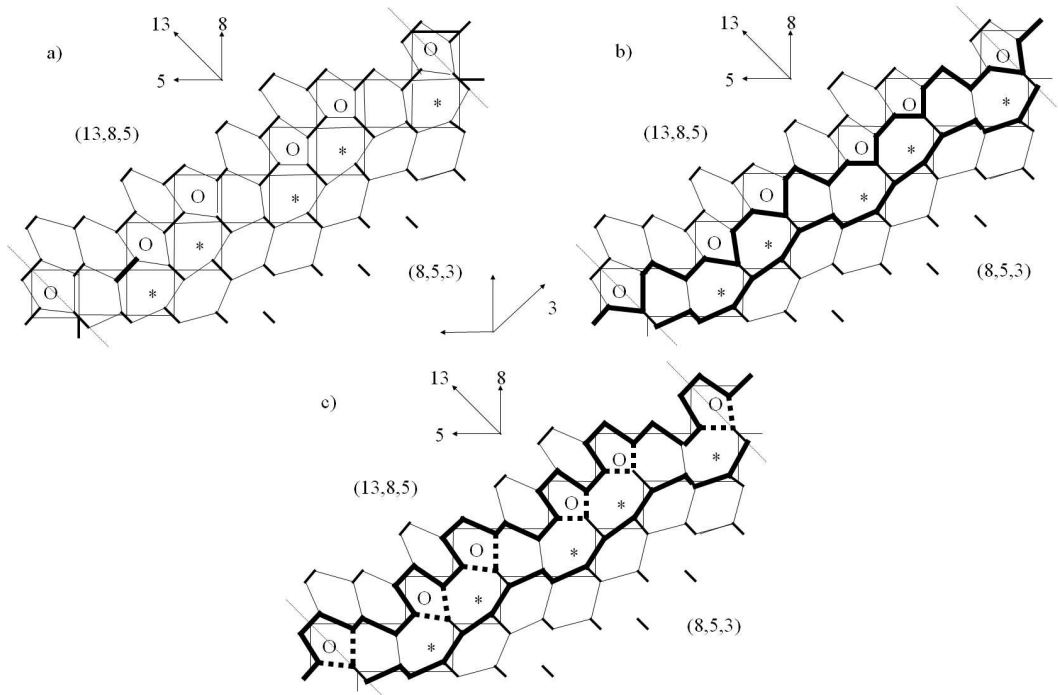}
}

\caption{Grain boundaries, layers and defect inclusions. a) Image of grain boundary (strip cut) on a square lattice (see also \cite{sadoc2}), with the geometrical shape of the cells outlined (exaggerated). The main parastichies $8$ (vertical, increasing northwards) and $5$ (horizontal, increasing westwards) are perpendicular. This is the boundary between grains $(13,8,5)$ (north-west) and $(8,5,3)$ (south-east).
It reads as a sequence $c(s) = [7,7,7,7,7,6,6,6,5,5,5,5,5]$ of successive cells on the generative spiral $s = 1,2,...$. The pentagons are labelled by O, the heptagons by $\star$, and the truncation of the vertices of the square cells is made accordingly. Flipping ($T1$) the thicker truncation would shift the grain boundary.
Since the main parastichies are orthogonal, the neighbor of cell $s$ is $s+8 {~}(mod~ 13)$ clockwise and $s+5 {~} (mod~ 13)$ anti-clockwise.
b) The thick lines delimit layer $\mu$.
%%%%%%%%%%%
Going up (north-westwards) from grain (8,5,3). Eight cells. The five pentagons of the grain boundary belong to layer $\mu+1$.
c) Going down (south-eastwards) from grain (13,8,5). Thirteen cells, but the five pentagons (O) are defect inclusions (bounded by the dotted lines): although neighbors of layer $\mu+1$ (by definition),
they are not neighbors of cells of layer $\mu-1$  \cite{oguey}.
%%%%%%%%%%%%%
}
\label{fg4}
\end{figure}
%______________________________________________________

\subsection{Grain boundaries as layers}

With this terminology, a grain boundary in phyllotaxis, e.g. $[7,7,7,7,7,6,6,6,5,5,5,5,5]$,  constitutes a complete layer $\mu$ going down, (south-eastwards), with its five pentagons as inclusions (Fig.\ref{fg4}c). Going up (north-westwards), only the five heptagons and the three hexagons constitute layer $\mu$, whereas the five pentagons are part of layer $\mu+1$. (Fig.\ref{fg4}b).

In phyllotaxis, layers have normally a Fibonacci number of cells. It is indeed the case of grain boundaries.
We shall see that the layers are invariant (a statement to be qualified) with respect to the choice of the first point, and thus under disappearance of the first cell. They fill up naturally under growth (addition of further cells).
Going down (or inwards in Fig.\ref{fg2}), a circular ring of defects constitutes a grain boundary, a complete layer of $f_{i+1}$ cells that are truncated squares, namely $f_{i-1}$ heptagons, $f_{i-2}$ topological hexagons and $f_{i-1}$ pentagons, successively along the $f_{i}$ turns of the generative spiral.

Layers are defined going down (inwards) from each grain boundary. The innermost layer of each grain may be incomplete. Notably, in the core (from the innermost ring of defect), one denotes the first layer $\mu = 1$ within parentheses () in the sequence $c(s) = (, , ,), \dots$ of cells along the generative spiral, characterized by their edge numbers.

\begin{figure}[tbp]
\resizebox{0.9\textwidth}{!}{%
\includegraphics{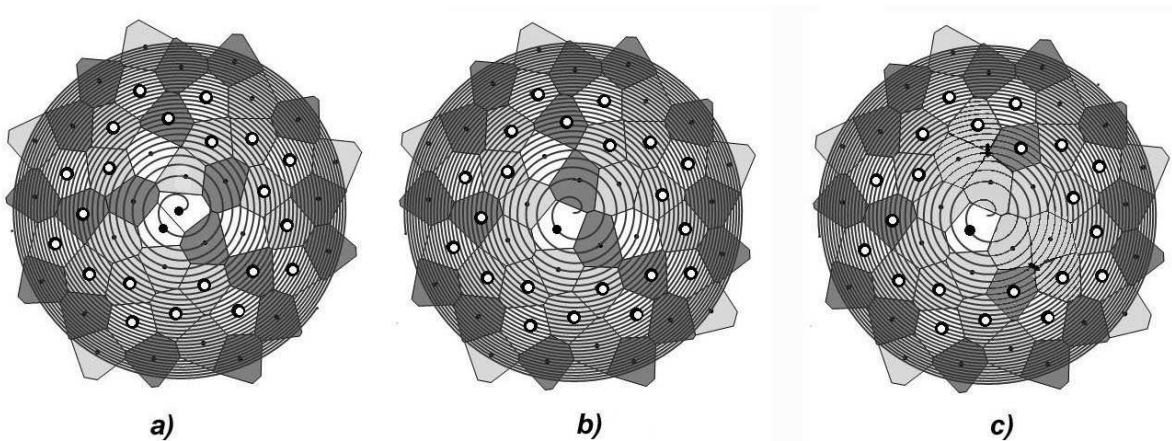}
}
\caption{Generative spiral, core and innermost layers of the phyllotactic structure. Detail of the center of Fig.2, $s \leq 50$, featuring the generative spiral, not directly visible in the structure. Here, dark, medium grey and white cells are respectively heptagons ($\star$), hexagons and pentagons ($(O)$) that abound in the core (the topological charge is $6$). Open circles label the innermost defect ring (twenty-one cells $s =10 - 30$); moving inwards, they constitute layer $3$ with the eight pentagons $s= 23-30$ as  defect inclusions (Fig.\ref{fg4}c).  A normal grain boundary would have all the eight cells $s=10-17$ heptagonal. Large black points mark the cells in layer 1. Small black points those of layer 2.
a) The point $s= 0$ is at the origin of the generative spiral, as in Fig.2.
Layer 2 consists of the eight cells $s= (4\star,7(O),2,5\star,8(O),3\star,6,9)$, cyclically. The two pentagons $s=7,8$ are defect inclusions. Layer 1 has only the two pentagons $s=0,1$.  Note that only cell $s=0$ has one non-Fibonacci neighbor $s=4$. b) Point $s= 0$ disappears, so that the first cell is $s=1$. This can be done topologically by removing the interface $0|2$. Layer 3 (the innermost defect ring) is unaffected. Layer 2 consists of the same eight cells
$s= (4,7(O),2\star,5,8(O),3\star,6,9)$, cyclically. Cells $s= 4,5$ are now hexagons and $s=2$ is now heptagonal. The two pentagons $s=7,8$ remain defect inclusions. Layer 1 has now only the pentagon $s=1$.
c) The core can be cleaned up by two $T1$ (thick interfaces) that delay the onset of parastichies $13$ from $s = 2,3$ to $s=4,5$ respectively.  Layer 3 (the innermost defect ring) is only affected in that $s=10,11$ are now heptagons and $s=15,16$ hexagons. Layer 2 consists of the same eight cells, now all hexagonal $s= (4,7,2,5,8,3,6,9)$, cyclically. The two cells $s=7,8$ remain defect inclusions, but they are now hexagons. Layer 1 has only the pentagon $s=1$.}

\label{newfig}
% %

\end{figure}

Figure \ref{newfig} gives the innermost layers of the planar phyllotactic structure of Fig.2, defined inwards from the innermost defect ring, the twenty-one cells $s =10 - 30$ constituting layer 3 (open circles)

% %______________________newfig__ Fig 6______________________________%

%________________________________________________________________ % % %
%

In spherical phyllotaxis, for $n=43-75$, the polar circle $[6,6,6,6,6,6,6,6,5,5,5,5,5]$ constitutes layer $\mu = 2$, going inwards, with the five pentagons $s=12-16$ as defect inclusions. Layer $\mu = 1$ ($(5,6,6)$) has the three cells $s=1,2,3$, whether going inwards or outwards. But, going outwards, layer $\mu=2$ has only eight cells, the five pentagons of the polar circle joining with the first eight hexagons of grain $ (13,8,5)$ to make up the thirteen cells of layer $\mu=3$.

For smaller spheres $n = 22-42$, the polar cap and circle $[(5,6,6),(5,5,5,5,5)]$ constitute the eight-cell layer $\mu = 1$ going inwards, with the five pentagons $s=4-8$ as defect inclusions.  Going outwards,  layer $\mu = 1$ ($(5,6,6)$) has only the three cells $s=1,2,3$, the five pentagons of the polar circle joining with the first three hexagons of grain $ (8,5,3)$ to make up the eight cells of layer $\mu=2$.

Now that the phyllotactic structure is understood in its details, we can discuss the action of an elementary topological transformation applied at vulnerable spots (a $T1$ on a grain boundary or the disappearance of the first cell). It turns out that the structure remains invariant under these specific transformations.

\section{Core, disappearance of the first cell and invariance of the phyllotactic structure}

In  equation (1),
the function $\rho(s)$ specifies amongst others the area  per cell (parameter $a$) and also the position of the first point ($s=0$  or $s=1$) with respect to the spiral's origin $\rho=0$, to be discussed now. The radial law of the generative spiral follows from demanding florets of equal sizes.

We denote by the operator $\bf D$, the action of an elementary topological transformation (whether a $T1$ or the disappearance of a cell) on the sequence $c(s)$ of cells $s=1,2,...$ along the generative spiral. We shall see that the phyllotactic structure (grain boundaries and organization into parastichies) is invariant under disappearance of the first cell $s=1$ of the sequence and suitable $T1$.

\subsection{Disappearance of the first cell $s=1$}

The first cell $s=1$ is a pentagon, with five neighbors $s = (2,3,6,9,4)$, in cyclic order (see Fig.1b), marking the onset of the parastichies $1,2,5,8,3$, that are the first Fibonacci numbers characteristic of phyllotaxis
%______________________ footnote5 ________
\footnote{In most cases (Euclidean phyllotaxis going outwards without a first point $s=0$ spot on the origin, or on the sphere ($16\le n \le 75$)), the first layer $s = (1,2,3)$ has three cells $(5,6,6)$. The second layer has eight cells $s = (4,7,10,5,8,11,6,9,4...)$ in cyclic order, with cell $s=4$ fitting between $s=1,2$, $s=5$ between $s=2,3$, $s=6$ between $s=3,1$, etc.. The cyclic order is obtained naturally by representing the circular layer of eight cell as a strip cut in a square lattice, as in the inset of Fig.\ref{f6}a, with the main, perpendicular parastichies $5$ (vertical, increasing northwards) and $3$ (horizontal, increasing westwards).}.
%_____________________________________________________________

What happens if cell $s=1$ of the tessellation disappears ($o$)? One sees (\cite{dubertret} or Fig.1b), e.g. by removing the boundary between cells $1$ and $4$, that cells $2$ et $9$ lose a side ($-$) and cell $4$, filling the space left vacant by $1$, gains a side ($+$).  All the other cells are topologically unchanged ($.$). Thus, disappearance of $s=1$ affects the sequence $c(s)$ of cells along the generative spiral through the operator ${\bf D} = (o,-,.,+,.,.,.,.,-,.,...)$   and the first cell is now $s=2$. The sequence
$[(5,6,6),5,5,5,5,5],6,6,6\ldots$
is {\it invariant} under disappearance of $s=1$.  Similarly, the sequence
$(5,6,6),[(6,6,6,6,6),(6,6,6),(5,5,5,5,5)],6,6,\ldots$
(spheres $n=43-75$), invariant under disappearance of $1$, with a $T1$ on $s=4$, as we shall see.

\subsection{A suitable $T1$ shifts a grain boundary}

In general, a grain boundary is shifted by a $T1$ at $s=p$, where $p$ labels the first heptagonal cell (Fig.\ref{fg4}a).
For example, the  $T1$ (${\bf D} =...,.[-,.,.,.,.,+,.,.,+,.,.,.,.],-,.,...$) translates the sequence
$...,6,[7,7,7,7,7,6,6,6,5,5,5,5,5],6,6,...$ into $...,6,6,[7,7,7,7,7,6,6,6,5,5,5,5,5],6,...$.
Remember that this is easily done because neighboring cells are close to squares, meeting at  critical vertices for $T1$ and ready to flip either way.

Thus, the sequence
$(5,6,6),[(6,6,6,6,6),(6,6,6),(5,5,5,5,5)],6,6,\ldots$
(spheres $n=43-75$) is invariant under disappearance of $1$, together with a $T1$ on $s=4$ displacing the first grain boundary
$[6,6,6,6,6,6,6,6,5,5,5,5,5]$
%_____________________ footnote6 ________
\footnote{ ${\bf D} = (o,-,.,[+,.,.,.,.,-,.,.,.,.,.,.,.],.,.,) \oplus (.,.,.,[-,.,.,.,.,+,.,.,+,.,.,.,.],-,.,) \\
= (o,-,.,[.,.,.,.,.,.,.,.,+,.,.,.,.],-,.,)$.}.
%_____________________________________________________________
Note that this thirteen-cell layer is too small to be a normal grain boundary (where the five innermost hexagons should be heptagons) but it has the correct topological charge of $+5$ that, with pentagon $s=1$, adds up to the charge $+6$ of an hemisphere. Let us call it a polar circle.

Other examples of sequences $c(s)$ are given in Appendix 1.

The sequence $c(s)$ and the local organization of cells into parastichies and grain boundaries are therefore invariant under disappearance of the first cell, or under growth, the inverse operation. This is a very important result that shows the malleability of the structure, capable (through $T1$ flips at grain boundaries) of responding easily to (shear) stress, whether local or global.

\subsection{Summary}

In summary, cells are organized in grains, concentric blocks separated by grain boundaries, surrounding a central core: A succession of large grains of hexagonal cells  that are concentric circular annuli, bounded and separated by circular grain boundaries $[f_{u-1}, f_{u-2}, f_{u-1}]$ made of $f_{u-1}$ heptagonal cells, $f_{u-2}$ hexagonal cells, and $f_{u-1}$ pentagonal cells.

Growth of the structure (i.e. addition of further cells $n+1,...$) can be achieved by addition of a first cell, under which the sequence $c(s)$ is invariant. The spiral lattice includes defects and grain boundaries that are responsive to local shear stress. Local chirality in the pattern gives the sense of shear and makes the response unambiguous.

Using foam terminology, a grain boundary in phyllotaxis constitutes one layer, with successive Fibonacci numbers of cells, depending on whether one counts outwards or inwards \cite{oguey}, with the difference - another Fibonacci number - made of (pentagonal) defect inclusions. In phyllotaxis (and not in foam), it is at a critical point and has a definite chirality, ready to flip under shear.

The grain boundaries serve as natural boundaries for optimal packing (one grain boundary constitutes a perfect circular boundary for the domain into which objects are to be packed \cite{sadoc1}). Outwards packing begins with the first complete grain boundary $[13, 8, 13]$ with 13 heptagons, 8 hexagons and 13 pentagons. The core is bounded by the 8 pentagons of the (first) incomplete grain boundary $[3, 5, 8]$. It has three heptagons instead of the full eight. The additional pentagons (nearly) fulfill the topological requirement that a tiled circular domain should have a topological charge of six \cite{rivier3,rivier2}.

The tiling is a foam made of topological hexagons, pentagons and heptagons, joined generically at vertices of degree three. Non-generic vertices appear on the grain boundaries. There is local organization throughout: The spirals (whorls) of first neighbors around each point (parastichies) are labelled by successive Fibonacci numbers \cite{sadoc2}.

But, unlike a foam, it is organized and encoded by a one-dimensional encoding of a spiral lattice \cite{koch1,coxeter,rothen}; this encoding is the sequence $c(s)$ of the number of sides $\{5,6,7\}$ of the Voronoi cells around successive points on the generative spiral. Consecutive cells along the generative spiral (except the first two) are not neighbors. The sequence $c(s)$ shows some structural patterns (grain boundaries) but conceals the local phyllotactic order of neighbor sites along the parastichies. In these sequences, we have bracketed  the grain boundaries as $[...]$ and gathered sites in groups $[(f_{i-1}), (f_{i-2}), (f_{i-1})]$, unless it is obvious.

We now discuss a dramatic manifestation of the organization of phyllotaxis and of its malleability in the agave.

% %______________________Fig.6________________________________ %
\begin{figure}[tbp]
\resizebox{0.9\textwidth}{!}{%
\includegraphics{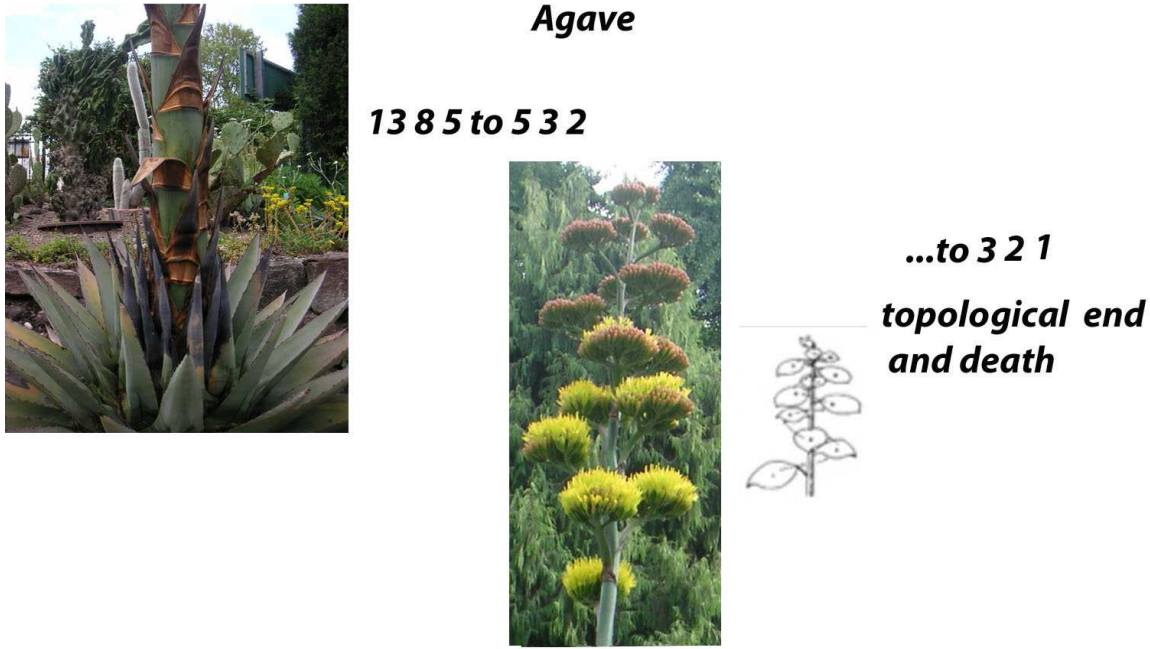}
}

\caption{ Agave Parryi. Structurally, it spends almost its entire life (24 years, approx.) as a single grain $(13,8,5)$ spherical phyllotaxis, a conventional cactus of radius $0.3$ m. During the last six month of its life, it sprouts (through three grain boundaries) a huge (2.5 m) mast terminating as seeds-loaded branches arranged in the (3,2,1) phyllotaxis, the final topological state before physical death. }
\label{f5}
% %

\end{figure}
%________________________________________________________________ % % %
% %______________________Fig.6 will be fig.7______________________________ %
\begin{figure}[tbp]
\resizebox{0.9\textwidth}{!}{%
\includegraphics{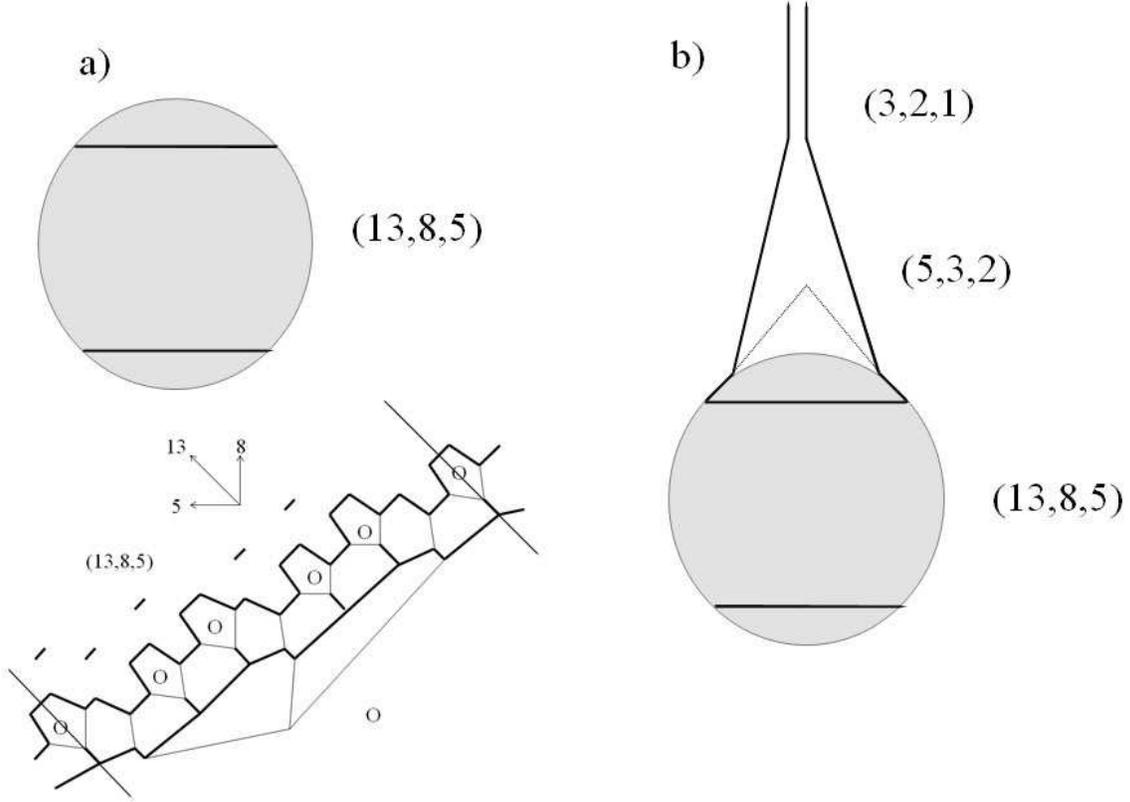}
}

\caption{ a) The agave during most of its life vegetates as a single grain $(13,8,5)$ spherical phyllotaxis. The north polar cap termination is given. b) Sprouting a mast avoids capping. A full grain boundary replaces the polar circle and
decreases the positive gaussian curvature of the sphere to zero, that  of the cone. }
\label{f6}
% %

\end{figure}
%________________________________________________________________ % % %

% %______________________Fig. 8______________________________ %
\begin{figure}[tbp]
\resizebox{0.9\textwidth}{!}{%
\includegraphics{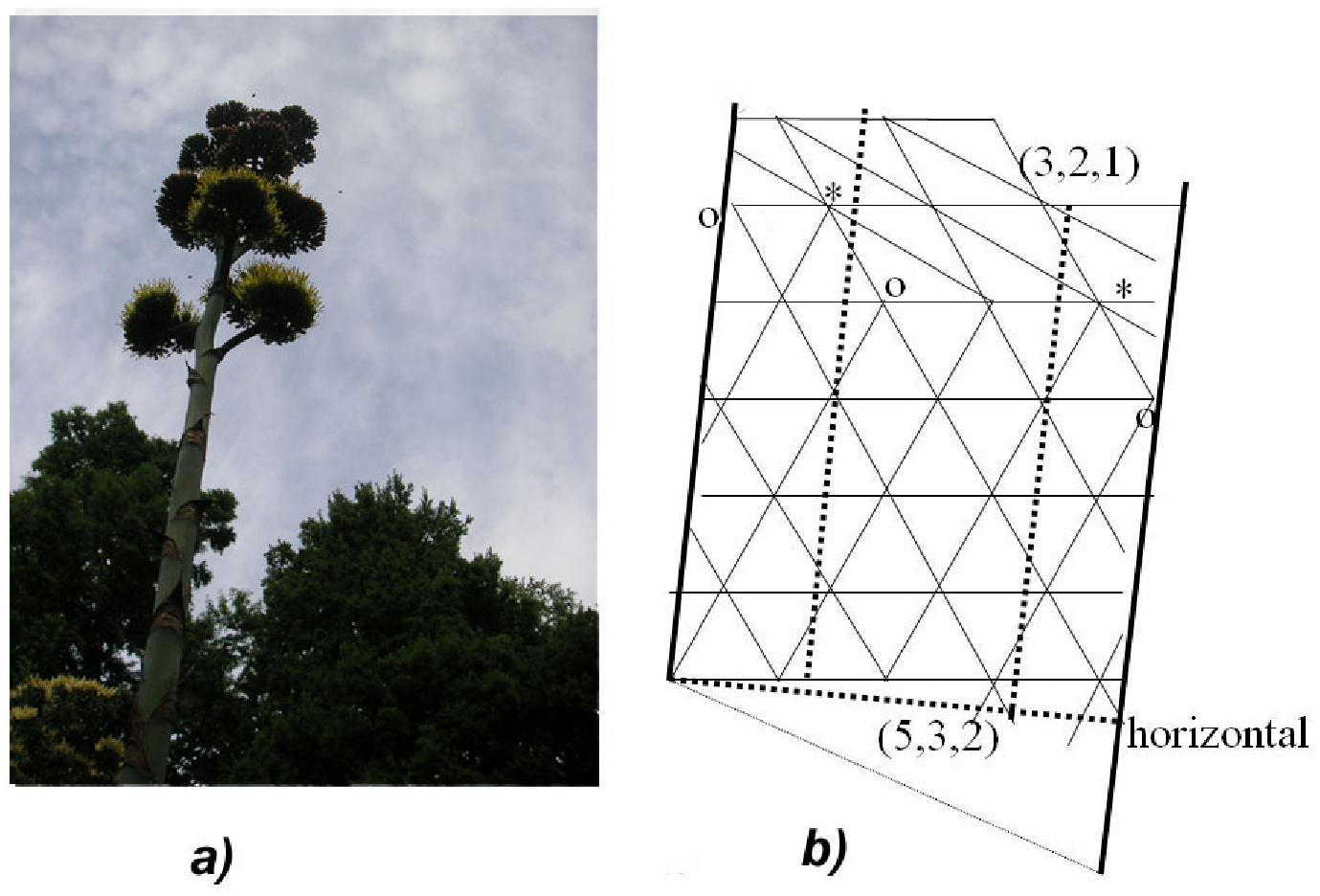}
}

\caption{  The mast of agave Parryi, seen in a photograph a) and as a strip representation tiled with equilateral triangles b).  The cylindrical mast is cut and flattened into a rectangular strip with the longer sides identified. The cylindrical mast is cut and flattened into a rectangular strip with the longer sides identified.  The mast is in the $(5,3,2)$ phyllotaxis, whereas the seeds-loaded branches are arranged in the $(3,2,1)$ phyllotaxis. The scales in the mast, or the seeds-loaded branches are the vertices of the triangular tiling. The narrower rectangle bounded by two dotted lines indicates the visible side of the mast in a). Note the grain boundary [(5,5),(6),(7,7)] between mast $(5,3,2)$ and seeds-loaded branches $(3,2,1)$. O, $\star$ represent vertices of degree 5 (pentagonal cells), respectively 7 (heptagonal cells). The lowest seeds-loaded branch marks the end of one of the five $5$ parastichies, replaced by the $1$ parastichy that begins there}
\label{f7}
% %

\end{figure}

\section{Anomalous growth: The agave}

An application of phyllotaxis to growth with the plasticity imparted by the grain boundaries
can be seen in the natural history of Agave Parryi (Fig.\ref{f5}). Structurally, it spends almost its entire life (24.5 years) as a conventional, spherical cactus in phyllotaxis $(13,8,5)$ which is also that of the pineapple.
During the last six month of its life, it sprouts a huge mast terminating as seeds-loaded branches.

The agave is a spherical phyllotaxis $n = 43-75$, $(5,6,6),[(6,6,6,6,6),(6,6,6),(5,5,5,5,5)],6,6,..;\dots$,
a single grain $(13,8,5)$ bounded by two grain boundaries (polar circles) $[(6,6,6,6,6),(6,6,6),(5,5,5,5,5)]$ and \\ $[(5,5,5,5,5),(6,6,6),(6,6,6,6,6)]$ on which the cells are square and the vertices critical points for $T1$ (Fig.\ref{f6}a).
Growth proceeds from the bottom, from the south to the north pole of the sphere. The south polar cap consists of a first layer of three cells $(5,6,6)$ with the south polar circle.

How does growth proceed beyond the north polar circle?  Normal growth (steady state) clings to the sphere. But in the last six months of its life, the agave first switches  to the cone tangent to the sphere on the north polar circle, through a complete grain boundary $[5,5,5,5,5,6,6,6,7,7,7,7,7]$ (a slight increase of the rate of growth replaces the five innermost hexagons by heptagons and the gaussian curvature falls from $+1/R^2$ to zero) (Fig.\ref{f6}b). Then, immediately, the agave switches to a less open cone, the mast in phyllotaxis $(5,3,2)$, through another smaller grain boundary. The mast terminates as seeds-loaded branches arranged on an open, uncapped cylinder in the $(3,2,1)$ phyllotaxis, through the grain boundary [5,5,6,7,7] (Fig.\ref{f7}). The cylinder in the $(3,2,1)$ phyllotaxis is indeed the topological endgame,  and, unlike a cone, it can be extended without end. If a ``Big Bang'' had occurred on the south polar cap, growth of a conical mast and a cylinder are the simplest means of avoiding a ``Big Crunch'' on the north polar cap.

Projection on the tangent cone (as in the mast of the agave) is different from the projection on the sphere because the generative spiral on the cone cannot have constant pitch $h(s)$ and keep a constant density of points. The pitch $h(s)$ must vary linearly with $s$. Also, the tangent cone is not capped, in fact, it can only end as a cylinder of seeds-loaded branches.

Thus, as summarized in one sequence $c(s)$ of grains and grain boundaries $[,,]$,\\
$..., 6,6,[5,5,5,5,5,6,6,6,7,7,7,7,7],[5,5,5,6,6,7,7,7],6,6,...6,[5,5,6,7,7],6,6,6...$.\\
Successive grains are the spherical phyllotaxis $(13,8,5)$, the mast $(5,3,2)$ on a cone and the seeds-loaded branches $(3,2,1)$ on a cylinder.

\section{Conclusions}
We have shown that the phyllotactic organization into parastichies and grain boundaries is robust. It survives cell disappearance and persists through growth. It accommodates shear stresses naturally. This is indeed the essential property of soft matter in general, and of foam in particular. Here the foam is not random, but organized into layers \cite{oguey}. A grain boundary is singularly effective as it is a layer of square cells joining at critical four-corner vertices, ready to flip under shear one way or the other. The disappearance of the first pentagonal cell that leaves invariant the organization into parastichies, is the other crucial agent of the malleability of the structure.

This paper has been written as a tribute to Professor Dominique Langevin  on her official retirement.
NR is grateful to F. Rothen and A.-J. Koch for many discussions and for their hospitality in Lausanne.

\section{Appendix 1: Examples of sequences $c(s)$}
In larger spheres ($n \ge 76$), a ring of defects straddling the equator, and palindromic, i.e. symmetric about it, emerges besides the two polar circles. For $n \ge 81$, this new ring splits into two mirror-imaged, normal grain boundaries with twenty-one cells each $[7,7,7,7,7,7,7,7,6,6,6,6,6,5,5,5,5,5,5,5,5]$.
The polar circle $[(7,6,6,6,6),(5,6,6),(5,5,5,5,5)]$ has been pushed towards the origin (the first layer $(5,6)$ has now two cells only). It will ultimately disintegrate, and the new grain boundary will carry the topological charge $+5$.

Indeed, in planar phyllotaxis (Fig.2 and footnote 3), the sequence is\\
$5,5,6,7,7,7,6,5,5,6,[(6,6,6,6,6,7,7,7),(6,6,6,6,6),(5,5,5,5,5,5,5,5)]\ldots$,\\
no longer constrained by symmetry about the equator. After removing the boundary ($0|2$)
and making  (i) one, or (ii) two $T1$ so that parastichies $13$ do not start from the first
layer, one has the invariant sequences \\
(i) $5,6,7,6,6,6,6,5,6,[(7,6,6,6,6,6,7,7),(6,6,6,6,6),(5,5,5,5,5,5,5,5)]\ldots$,\\
or
(ii) $5,6,6,6,6,6,6,6,6,[(7,7,6,6,6,6,6,7),(6,6,6,6,6),(5,5,5,5,5,5,5,5)]\ldots$.
Then, all cells have neighbors on parastichies labelled by Fibonacci numbers. Sequence (i) is obtained if one shifts the first point from the origin of the generative spiral, i.e, set $s \in \emph{N} + 1/2$ in eq.1.
%_________________________________________________
\section{Appendix 2: Details of the core of the phyllotactic structure}

Consider Fig.\ref{exfig9}, which is Fig.\ref{newfig}a with the cells numbered. The core includes all the cells ($s=0-9$) inside the innermost defect ring (open circles, $s=10-30$). Moving inwards, the twenty-one cells of the innermost defect ring constitute layer 3 with the eight pentagons $s= 23-30$ as defect inclusions.  Layer 2 consists of the eight cells $s= (2,5(\star),8(O)),3(\star),6,9,4(\star),7(O),)$, cyclically. The two pentagons $s=7,8$ are defect inclusions. Layer 1 (thick black points) has only the two pentagons $s=0,1$.

The generative spiral gets tighter as one goes out (eq. (1), with $\rho \sim \sqrt{s}$) so that cells have the same average size on average.
The azimuthal angle between two successive points on the spiral $2\pi/\tau$  is a substantial fraction $(0.618)$ of $2 \pi$. As a consequence, successive cells on the generative spiral are not direct neighbors (except the first two).

Let us follow the generative spiral. Cell $s=0$ is at the origin. With cell $s=1$, it constitutes layer 1 (thick black points). The azimuth of cell $s=2$, the first cell of layer 2, is greater than $2\pi$, so that the generative spiral has made a complete turn by point $s=2$. The last cell $s=9$ of layer 2 has nearly the same azimuth as cell 1 - indeed, they are neighbors, belonging to one of the eight parastichies $8$ - and by then, the generative spiral has made five turns. Cell $10$ (the first cell of layer 3) is neighbor to cell $2$ and they belong to another parastichy $8$. Layer 3 (open white circles) with twenty-one cells $s=10-30$ uses up thirteen turns of the generative spiral, each cell containing thirteen segments. This is why the last eight cells $s=23-30$, which are pentagons, look grey (shaded) in Fig.\ref{exfig9}.

Further out, the phyllotactic structure is fully developed, with defect annuli separating hexagonal grains. For example, in Fig.2, the fourth ring of defects from the outside has $f_{10} = 55$ cells (21 heptagons, 13 hexagons and 21 pentagons). It separates grains $(34,21,13)$ and $(55,34,21)$.

\end{document}